# Review of criticisms of ballistic pressure wave experiments, the Strasbourg goat tests, and the Marshall and Sanow data

Michael Courtney, PhD
Ballistics Testing Group, P.O. Box 24, West Point, NY 10996
Michael_Courtney@alum.mit.edu

Amy Courtney, PhD
Department of Physics, United States Military Academy, West Point, NY 10996
Amy_Courtney@post.harvard.edu

*Abstract:*
This article reviews published criticisms of several ballistic pressure wave experiments authored by Suneson *et al.*, the Marshall and Sanow "one shot stop" data set, and the Strasbourg goat tests. These published criticisms contain numerous logical and rhetorical fallacies, are generally exaggerated, and fail to convincingly support the overly broad conclusions they contain.
*Originally submitted 13 December 2006. Revised version submitted 31 July 2007.*

## I. Introduction

Selecting service caliber handgun loads with the greatest potential for rapid incapacitation of violent criminal or terrorist attackers is of great interest in the law enforcement community [PAT89]. This interest has fueled heated debate regarding the scientific merits of some contributions.

There are many examples in the scientific literature addressing research of questioned quality. However, the use of *ad hominem* attack and other rhetorical fallacies is relatively rare in the peer-reviewed literature. One example where the scientific community is constantly striving to counter low-quality science is in the area of claimed health benefits of certain foods or nutritional supplements. Literature addressing this kind of pseudo-science shows greater care, respect, and restraint from using fallacies and exaggerations than the ballistics literature reviewed here.

This review of exaggerated criticisms is more explicit than many articles in reminding readers of the essential elements of scientific method and publication, because this article should be understood by the layman as well as the trained scientist.

## II. Position being defended in critical reviews

It is generally agreed that bullet design plays at least as important a role in bullet effectiveness as the cartridge from which it is fired. However, it is still widely debated whether the only contributing factors to the effectiveness of different loads are the volume of crushed tissue and penetration depth [PAT89]. Crushed tissue volume and penetration clearly contribute to bullet effectiveness through the physiological consequences of blood loss. Over the years, mechanisms more heavily dependent on energy transfer have been suggested such as hydrostatic shock, hydraulic reaction [CHA66], and the temporary stretch cavity. Authors who suggest these mechanisms usually have something in mind more or less related to a ballistic pressure wave.

The view that the crushed tissue volume (the permanent cavity) [PAT89, FAC87a, FAC96a, MAC94] is the only reliable contributor to incapacitation (for handgun bullet placements that do not hit the central nervous system or supporting bone structure) depends strongly on the unproven presupposition that easily detectable wounding[1] is necessary to contribute to incapacitation. Fackler summarizes the position that there are only two wounding mechanisms for all bullet injuries [FAC87a]:

*Tissue crush is responsible for what is commonly called the permanent cavity and tissue stretch is responsible for the so-called temporary cavity. These are the sole wounding mechanisms.*

Duncan MacPherson [MAC94 p63] gives the reasoning behind the assertion that temporary cavitation plays little or no role in incapacitation via handgun bullets:

*The short barrels of practical handguns limit the bullet velocities that can be reasonably achieved. Testing has shown that the diameter of handgun bullet temporary cavities is usually less than 10 centimeters (4 inches) even at depths near the surface that do not contain vital organs. The strain produced by these small cavities is below the elastic limit in most tissue, and so there is usually little or no damage. There are impact areas where temporary cavity produced by handgun bullets causes wound trauma incapacitation, but these areas are relatively small. Bullet impact in these atypical*

---

[1] By this we mean wounding typically observed by a trauma surgeon or medical examiner.

*tissues is by chance, not intent, and this is unlikely because these areas are a small fraction of the total. The temporary cavity produced by handgun bullets is usually not a factor in wound trauma incapacitation.*

Since these authors assert that there are only two projectile wounding mechanisms (permanent cavitation and temporary cavitation), and that handgun bullets do not create significant wounding via temporary cavitation, they conclude that the permanent cavity (crushed tissue volume) is the only wounding mechanism in handgun bullets. This is summarized by Patrick [PAT89]:

*The tissue disruption caused by a handgun bullet is limited to two mechanisms. The first, or crush mechanism is the hole the bullet makes passing through tissue. The second, or stretch mechanism is the temporary cavity formed by the tissues being driven outward in a radial direction away from the path of the bullet. Of the two, the crush mechanism, the result of penetration and permanent cavity, is the <u>only</u> handgun wounding mechanism which damages tissue.*

Most of the criticisms considered here are either authored by proponents of this view or published in a journal whose editor is a proponent of this view.

A recurring comment in articles expressing this view is that the pressure wave does not move or damage tissue [FAC87a, MAC94]. This is based on an experiment from 1947 [HKO47], as well as the observation that the ultrasonic shock waves of lithotripsy do not damage tissue.

Harvey *et al.* [HKO47] studied macroscopic effects of pressure waves on tissue, but did not consider neurological tissue, and tissue was not examined with any advanced microscopic or chemical technique needed to detect damage caused by pressure waves in neurological tissue. Reliance on Harvey *et al.* to support the view that pressure waves do not contribute to neurological wounding or incapacitation contains several errors:

- The presupposition that macroscopic movement is required for tissue damage.
- The presupposition that neurological wounding would have been observed in these experiments with the older techniques employed.
- The presupposition that easily observable wounding is necessary for incapacitation.

The assertion that lithotriptors prove that pressure waves do not damage tissue is often phrased something like [FAC96a]:

*A modern researcher wishing to study the sonic pressure wave could avoid the confounding effects of cavitation by using a lithotriptor. The lithotriptor generates sonic pressure waves without using a projectile, therefore yielding no temporary cavity . . . these sonic waves do not significantly harm the surrounding tissue.*

Not only does this suggested experimental design have the problem that ballistic pressure waves are much broader in frequency spectrum than lithotriptor waves, more recent research has revealed that the pressure waves associated with lithotripsy do indeed have the potential for significant tissue injury [EWL98, LOS01, LKK03].

Proponents of the view that wounding and incapacitation depend only on permanent and temporary cavitation have done considerable research validating the performance of ballistic gelatin as a reliable tissue simulant [MAC94, FAC96a, WOL91] and have been vocal proponents of the idea that expanded bullet diameter and penetration (parameters easily measured in gelatin) are the only reliable contributors to incapacitation via handgun bullets. However, they have not published careful studies correlating observed metrics of incapacitation with easily observable wounding or these parameters easily measured in gelatin.

In defending this position in light of contrary research findings [SHS87, SHS90a, SHS90b, TCR82, MAS96, STR93] proponents have relied primarily on exaggerated criticisms of these research findings and personal attacks on the authors rather than publishing or referencing incapacitation data to support their position.

### III. Exaggerated criticisms of ballistic pressure wave experiments of Suneson *et al.*

Suneson *et al.* [SHS90a, SHS90b] report that peripheral high-energy missile hits to the thigh of anesthetized pigs cause pressure changes and damage to the central nervous system. This experimental study on pigs used high-speed pressure transducers implanted in the thigh, abdomen, neck, carotid artery, and brain [SHS90a p282]:

*A small transducer . . . mounted in the end of a polyethylene catheter . . . was implanted into the cerebral tissue in the left frontoparietal region about 10 mm from the midline and 5 mm beneath the brain surface through a drill hole (6-mm diameter).*

The sensor implanted in the brain measured pressure levels as high as 46 PSI and 50 PSI for pigs shot in the thigh as described. (See Figure 2C and 2D [SHS90a p284].) The average peak positive pressure to the brain over the different test shots with that set-up was 34.7 PSI +/- 9.7 PSI. The error range does not represent uncertainty in individual measurements, but rather uncertainty in determination of the mean because of significant shot-to-shot variations in the pressure magnitude reaching the brain. (For a given local pressure wave magnitude in the thigh, the distant pressure wave magnitude in the brain will show variation.)

Apneic (non-breathing) periods were observed during the first few seconds after the shot, and both blood-brain and blood-nerve barrier damage were found. They concluded "distant effects, likely to be caused by the oscillating high-frequency pressure waves, appear in the central nervous system after a high-energy missile extremity impact."

Fackler published a reply [FAC91a, see also FAC89a], asserting that "Shock Wave" is a myth:

*In ascribing "local, regional, and distant injuries" to the sonic pressure wave, Suneson et al. have overlooked the effect of transmitted tissue movement from temporary cavitation. Since two distinct mechanisms are acting in the Suneson et.al experiment, one cannot arbitrarily assign any effects observed to only one of them.*

Fackler's reply has several major flaws:
- Fackler employs the straw man fallacy by referring to the pressure wave studied by Suneson *et al.* as "the sonic pressure wave." The authors studied a "shock" wave and stated that the wave includes ultrasonic frequency components up to 250 kHz. In addition, Fackler considers the "sonic pressure wave" to be limited to a very short (several microseconds) pulse that precedes temporary cavitation. Suneson *et al.* are describing effects of pressure waves with a longer duration (roughly 1 ms).
- Fackler creates a false dichotomy to divide effects beyond the permanent crush cavity into only the temporary cavity and the "sonic" pressure wave. The pressure wave consists of every force per unit area that can be detected by a high-speed pressure sensor. The leading edge of the pressure wave travels at sonic velocity, but the pressure wave also has components that propagate more slowly.
- Tissue movement by cavitation is not distinct from the ballistic pressure wave. (One can consider temporary cavitation an effect of the inertial component of the pressure wave [FAC96a].) Consequently, ascribing the local neural injuries to the pressure wave is not unreasonable, though Fackler is correct to point out that in the local region, the pressure wave effects cannot be distinguished from temporary cavitation effects.
- Suneson *et al.* also report regional and distant effects beyond the reach of the temporary cavity. Nerve damage is observed as far as 0.5 m away from the wound channel. These regional and distant effects cannot be ascribed to temporary cavitation.

Fackler continues:

*Recently, eleven adult human-sized swine (90 kg) were shot in the proximal part of the hind leg with a projectile producing the damage profile of the Russian AK-74 Assault rifle bullet. This same projectile was used in another study in which five 90 kg swine were shot through the abdomen …No indication of any sort of "distant" damage was seen in the pigs' behavior and no "distant" injuries were found at autopsy.*

The methodology of Fackler's pig experiments [FBC89] is significantly different from that of Suneson *et al.*, who report that the neural damage is not easily observable, but rather depends upon examination with light and electron microscopy. The effects that Suneson *et al.* report "were evident a few minutes after the trauma and persisted even 48 hr after the extremity injury." In Fackler's experiments, autopsies were not performed until weeks or months later. With such great differences in experimental methodology, it is unfounded to assert that Fackler's swine experiments contradict the conclusions of Suneson *et al.*

Fackler continues:

*A review of 1400 rifle wounds from Vietnam (Wound Data and Munitions Effectiveness Team) should lay to rest the myth of "distant" injuries. In that study, there were no cases of bones being broken, or major vessels torn, that were not hit by the penetrating bullet.*

It is unreasonable to refute modern observations (using new methods) of microscopic damage to nerve cells by referring to the absence of observations of broken bones or torn blood vessels in Vietnam-era observations from trauma surgeons. The Vietnam-era study was not looking for distant nerve damage and did not employ the methods used by Suneson *et al.*

Distant injuries are not a "myth" as Fackler claims. Suneson *et al.* finds agreement with later experiments in dogs conducted by an independent research group using a similar method [WWZ04]:

*The most prominent ultrastructural changes observed at 8 hours after impact were myelin deformation, axoplasmic shrinkage, microtubular diminution, and reactive changes of large neurons in the high-speed trauma group. These findings correspond well to the results of Suneson et al., and confirmed that the distant effect exists in the central nervous system after a high-energy missile impact to an extremity. A high-frequency oscillating pressure wave with large amplitude and short duration was found in the brain after the extremity impact of a high-energy missile . . .*

This experiment in dogs is not the only evidence in the literature tending to confirm the findings of Suneson *et*

*al.* The lateral fluid percussion model of traumatic brain injury has demonstrated conclusively that transient pressure pulses in the same magnitude range as observed by Suneson *et al.* can cause traumatic brain injury and incapacitation [THG97, TLM05, and references therein].

**IV. Exaggerated criticisms of the Strasbourg goat tests**

The Strasbourg Tests [STR93] studied handgun bullet effectiveness in goats by shooting the animals broadside through the center of the chest and recording the time to incapacitation (falling down). The published data includes incapacitation times from shooting 580 goats with 116 different handgun loads.

These tests employed a pressure sensor inserted into the carotid artery of live unanaesthetized goats. These tests directly suggest that an internal pressure wave created by the interaction of the bullet and tissue can contribute to rapid incapacitation and can incapacitate more quickly than the crush cavity/blood loss mechanism alone [STR93]:

*In a substantial number of cases, the subject was incapacitated almost instantly. Each time this occurred, between two and five pressure spike tracings of high amplitude and short duration were found which immediately preceded and matched corresponding, diffused, or flattened lines (EEG tracings). Normally, the time lag between the first pressure spike and the beginning of slowed or flattened lines was between 30 and 40 milliseconds (although there were several cases where this delay lasted as long as 80 milliseconds)…The taller pressure spike tracings always preceded the slowed or flat line tracing…The initial spikes had to be of a certain height in order for the animal to collapse immediately.*

The average incapacitation times show good correlation (R =0.91) with an empirical model based on pressure wave magnitude [COC06c]. The relationship between pressure wave and incapacitation times in the Strasbourg tests[2] are consistent with the Suneson *et al.* observations [SHS90a, SHS90b, SHK90, SHL89, SHS88, SHS87], the Wang *et al.* observations in dogs [WWZ04], observations of incapacitation and traumatic brain injury in fluid percussion model research [THG97, TLM05, and references therein], as well as an incapacitation study in deer [COC06d].

However, the Strasbourg tests have been severely criticized on a number of points [FAC94a, FAC97a].

*A. Ad hominem and bandwagon fallacies*

Usually a concluding paragraph in a scientific journal summarizes the main points of the article, draws conclusions, or suggests areas of possible future research. Outside of the critical works referenced and examined here, debate in the peer-reviewed journals is almost completely free of *ad hominem* attacks, appeal to ridicule, and the bandwagon fallacy.

The conclusion of Fackler's criticism on the Strasbourg tests contains a curious combination of all three [FAC94a]:

*The only people who think the "Strasbourg Tests" are real are the usual crowd of crackpot "magic bullet" believers and the pathetically incompetent editors of consumer gun magazines like Guns & Ammo. I suppose we'll soon see anonymous reports "proving" that Elvis is alive and conducting one shot stop experiments on unicorns. And, of course, someone will believe that too.*

This kind of attack and fallacious reasoning is expected in political ad campaigns and internet discussion groups, but it is shocking to find it in the peer-reviewed literature.

If the report is a fraud, attempts to repeat the experiment would obtain different results (significantly different average incapacitation times, for example) than those reported. In the absence of conclusive eyewitness, documentary, or physical evidence of fraud, failure to repeat experimental results under the same conditions is the only sound scientific basis for a definitive conclusion that results are fraudulent.[3]

*B. Fallacy of expert opinion/appeal to authority*

It is untrue that expert opinion rather than repeatable experiments is the ultimate arbiter of scientific truth. The Copernican model of a heliocentric solar system was right, even though the experts of his day disagreed. The theory that DNA contains the inherited genetic information was not immediately accepted by the experts of the day. It was right because it was ultimately confirmed by repeatable experiments. The theory of cold fusion wasn't doubted merely because expert opinion held it to be impossible. On the contrary, attempts to repeat it have failed.

In the absence of attempts to repeat an experiment, the best that expert opinion and peer review can offer is an early assessment on the likely validity and value of a new scientific claim or result. Ultimate validity and value can only be determined in the light of work repeating the original, whether precisely or in ways that attempt to confirm the same principles as the original.

---

[2] More detailed time-domain analysis also reveals a fast incapacitation mechanism that correlates well with peak pressure magnitude.

[3] And even then, many researchers would be given the benefit of the doubt and the discrepancy would tend to be considered as uncontrolled conditions, an invisible factor impacting the outcome, or an unexplained glitch. Most scientists are slow to reach the conclusion of fraud. However, combined with anonymity, failure to repeat an experiment would probably be considered sufficient grounds for fraud.

Yet in Fackler's review of Strasbourg, the initial opinion of a small group of experts is taken to establish with finality (see the conclusion quoted above) the fraudulent nature of the report, and this without eyewitness, documentary, or physical evidence suggesting fraud [FAC94a]:

*The FBI committee, which includes a half dozen of the world's most highly regarded gunshot-expert forensic pathologists, felt that the organization and wording of the document betrayed it as a hoax. Why else would experimental results be circulated anonymously?*

*Reputable scientists put their names on their work, take responsibility for it and respond to critical reviews by their colleagues.*

The matter of anonymous authorship will be addressed separately. However, if full disclosure of contributing scientist identity is so important to the validity of a work, one wonders why the author fails to name these six "most highly respected gunshot-expert forensic pathologists."

If "review of a work by other recognized scientists is mandatory before any results are accepted as valid" then why do six contributors of the opinion that the Strasbourg tests are a hoax remain anonymous? The FBI committee remains anonymous while stating anonymity as the criterion used to determine that the Strasbourg report is a hoax.

In addition, Fackler's appeal to authority is fallacious because it fails to meet the several widely accepted standards for appeal to authority. [4]

- *The claim being made by the person is within their area(s) of expertise.*

*If a person makes a claim about some subject outside of his area(s) of expertise, then the person is not an expert in that context. Hence, the claim in question is not backed by the required degree of expertise and is not reliable.*

The science of ballistic incapacitation combines physics, wounding, neurology, physiology, and behavior. No traditionally trained doctor or scientist qualifies as an authoritative expert to the degree that their opinion is believable without reference to repeatable experiments to support their point.

Forensic pathologists are trained and experienced in determining the cause of death that is not necessarily the same as the cause of incapacitation. Bullets that do not hit the CNS often kill by blood loss, but this takes at least several minutes. A forensic pathologist almost never performs the sensitive microscopic examination required to detect diffuse axonal injury or chemical tests to determine whether remote injury to the brain might have caused incapacitation prior to death.

In the absence of authoritative experts in the science of ballistic incapacitation, sound arguments must always refer to repeatable experiments rather than expert opinion/appeal to authority.

Another requirement for a valid appeal to authority is:

- *There is an adequate degree of agreement among the other experts in the subject in question.*

*If there is a significant amount of legitimate dispute among the experts within a subject, then it will be fallacious to make an Appeal to Authority using the disputed experts. This is because for almost any claim being made and "supported" by one expert, there is a counterclaim that is made and "supported" by another expert. In such cases an Appeal to Authority would tend to be futile. In such cases, the dispute has to be settled by consideration of the actual issues under dispute.*

Ongoing disputes in the field of wound ballistics are obvious in much of the literature reviewed here as well as Fackler's own review articles where he presents his side of the debate [FAC96a, FAC88a].

A third requirement violated by Fackler's fallacious appeal to authority is:

- *The authority in question must be identified.*

*A common variation of the Appeal to Authority fallacy is an Appeal to an Unnamed Authority . . .*

*This fallacy is committed when a person asserts that a claim is true because an expert or authority makes the claim and the person does not actually identify the expert. Since the expert is not named or identified, there is no way to tell if the person is actually an expert. Unless the person is identified and has his expertise established, there is no reason to accept the claim.*

Since the six forensic pathologists were not identified, there is no reason to accept their opinion that the Strasbourg report is a hoax.

---

[4] *(Italicized material on fallacious appeal to authority quoted from www.nizkor.org/fallacies/appeal-to-authority.html, accessed 11/4/2006).*

It is difficult to assess whether a fourth requirement for a valid appeal to authority is met:

- *The person in question is not significantly biased.*

*If an expert is significantly biased then the claims he makes within his area of bias will be less reliable. Since a biased expert will not be reliable, an Argument from Authority based on a biased expert will be fallacious. This is because the evidence will not justify accepting the claim.*

Since Fackler's appeal to authority based on the opinion of six unnamed forensic pathologists fails to meet at least three (possibly four) of the required criteria for valid appeals to authority, this appeal to authority is fallacious.

### C. Criticisms of experimental design

*i. Goats are not humans*

Variations of this criticism appear with some regularity, and these criticisms are generally a valid caution against interpreting the results of bullet testing in goats directly for selection of handgun ammunition for self-defense against humans.

However, the value in the goat tests is not in the unwarranted presupposition that the loads that work best in goats will work well in humans. The value in the goat tests is the study of different incapacitation mechanisms under carefully controlled conditions. If a given incapacitation mechanism is shown to exist in goats, it might exist in humans also, opening the door to further work to establish links between humans and the animal test subjects.

*ii. Head shots destroy necessary evidence*

The Strasbourg test methods included euthanizing the goats by shooting them in the head. Fackler asserts that this is an experimental design error [FAC94a]:

*The authors of the purported tests postulated some mysterious pressure-mediated effect on the brain that causes "incapacitation." The critical part of any such study would be removal of the brain to see if it showed any physical evidence to account for the postulated effects. In the purported "study," after the animals were shot in the chest, and the time that they remained standing recorded, they were killed BY SHOOTING THEM IN THE HEAD. This would destroy any possibility of establishing the mysterious "incapacitation" postulate as fact.*

The fact that a ballistic pressure wave originating remotely reaches the brain and can cause neurological damage in the brain had already been established by Suneson *et al.* in papers published prior to the Strasbourg tests [SHS90a, SHS90b and references therein]. The results of microscopic brain examination using modern neurological techniques would have had some value, but it is not essential given the prior work of Suneson *et al.*

Suneson *et al.* linked pressure waves to neurological damage. The Strasbourg design probes the link between pressure wave and incapacitation without investigating detailed mechanisms of neurological damage. (Similarly, a study can link a hazardous chemical with cancer without investigating the biomolecular processes. Biomolecular details can be established independently.)

*iii. Pressure wave effect should be immediate*

Fackler writes:

*Even if we presume pressure on the brain from transmitted temporary cavitation via pressure pulses in the aorta and carotid vessels to be large enough to cause an effect on the goats, everything we know about cerebral physiology and pathology suggests that any such effect, if it did occur, would be IMMEDIATE.*

This is the accent fallacy and untrue. As discussed later, the pressure wave origin is not restricted to temporary cavitation compressing blood vessels. Furthermore, the kind of mild traumatic brain injury (TBI) thought to cause incapacitation is probably a mild (Grade III) concussion. The unconscious period associated with mild concussion is not always immediate [SHA02].

Fackler continues:

*In the purported "Strasbourg tests" none of the goats fell over . . . immediately; in most of the purported test shots the magic took from 5 to 40 seconds to work.*

The Strasbourg authors do not suggest a pressure wave effect in cases where the test subject takes a relatively long time to fall down. In contrast, the pressure spikes are correlated with occasions where the goats fall almost instantly [STR93]:

*In a substantial number of cases, the subject was incapacitated almost instantly. Each time this occurred, between two and five pressure spike tracings of high amplitude and short duration were found…*

The authors don't give a precise time range for "almost instantly." However, the data shows incapacitation times less than 5 seconds are only produced by the highest pressure waves delivered in the chest [COC06c]. Therefore, instances where the test subject falls in under 5 seconds give clearest evidence of a pressure wave mechanism. Cases where the test subject remains standing for 5-40 seconds are instances where the pressure wave reaching the brain has little or no impact on incapacitation. In other words, the blood loss mechanism seems dominant in these cases.

*iv. Pressure wave via temporary cavitation compressing aorta would be smaller in humans*

The objection that the pressure wave created by temporary cavitation compressing the aorta would usually be smaller in humans is true, but it ignores the other components of the pressure wave [FAC96a], including the high-frequency components that Suneson *et al.* had already shown to contribute to TBI [SHS90a, SHS90b, and references therein].

*v. Tests involving live unanesthetized animals would be prohibited in the USA*

There are exceptions to the anesthesia requirements in the National Academy of Sciences (NAS) guidelines for research conducted in a farm environment and for research in which withholding anesthesia is important to the scientific goals of the experiment. While wounding can be observed, quantified, and studied in test subjects under anesthesia, it is not clear that incapacitation can be. If there is reasonable support that anesthesia might have changed the outcome of the experiment, shooting unanesthetized goats would have been allowed.

There are also other exceptions in the NAS guidelines that allow for shooting unanesthetized animals. Examples are routine slaughter and hunting. Numerous gun writers have reported studies of bullet performance in wild game. The Thompson-LaGarde stockyard tests combined a bullet effectiveness study with livestock [LAG16][5].

*D. Criticisms of experimental results*

*i. Rib hits correlate too well with long incapacitation times*

The Strasbourg report contains the unlikely result that for 115 of 116 loads tested the case with the longest incapacitation time (in 5 shots) was one in which a rib was hit. Fackler assigns a fraudulent motive [FAC94a] to an experimental result that can more simply be ascribed to lack of objective criteria to determine whether or not a rib was hit.

Since the Strasbourg report does not describe criteria for determining whether a rib was hit, reservations about those assignments are justified. Having shot numerous similarly-sized deer ourselves [COC06d], it is challenging to say with certainty whether or not a bullet actually hit a rib or whether the rib was damaged by the pressure wave or temporary cavitation of a near miss. Bullets with larger pressure waves/cavitation almost always break at least one rib on entry in deer, even when placed between two ribs.

Imaging techniques might objectively determine whether a rib is damaged by a passing bullet but would not indicate whether or not the bullet hit the rib. It is not possible to objectively determine whether or not the bullet hit a rib. The error in Strasbourg is the subjective assignment of whether or not a rib was hit. Thus, the Strasbourg data set is reasonable for incapacitation times, but not regarding any analysis based on whether or not ribs were hit.

*ii. Invokes a "bizarre and heretofore unexplained phenomenon"*

The pressure wave incapacitation mechanism may have been "unproven" or "doubted" at the time of the Strasbourg report, but there is sufficient reference to it in both the scientific and popular ballistics literature that referring to it as "bizarre and heretofore unexplained" is an exaggeration.

Harvey *et al.* showed that bullets produce significant pressure waves [HKO47]. In tests conducted at the Aberdeen Proving Grounds from 1928 to 1930, Frank Chamberlain determined that "explosive effects" cause tissue destruction in all directions far beyond the wound channel [MAS96 p26-27]. Chamberlin favored the theory of "Hydraulic Reaction" of body fluids [CHA66] as an important factor in wounding. Goransson *et al.* reported EEG suppression in pigs that was attributed to a remote effect [GIK88], and Suneson *et al.* [SHS90a, SHS90b] had reported brain injury attributed to distant effects transmitted via pressure wave of a bullet striking the thigh in pigs. A study in humans [OBW94] later determined that remote pressure wave effects cause injury in humans.

Ascribing wounding and incapacitation to a pressure wave has a long history in wound ballistics. In addition to the references above, there are a number of articles in the literature [LDL45, PGM46, GIK88, MYR88, WES82, TCR82]. Fackler's knowledge of this history is apparent in articles asserting his point of view [FAC87a, FAC96a]. There is a difference between supporting one's own view in a scientific debate and pretending that the other point of view is a completely new idea.

*iii. "Near perfect" correlation with M&S OSS rating*

In a critical review of the second M&S book [MAS96], Fackler states [FAC97a]:

*... the near perfect correlation of Marshall's random torso "one-shot stops" with the purported goat shot results is strong evidence that the anonymous "Strasbourg Test" results have*

---
[5] This was before the NAS guidelines, but probably would have been allowed under the current rules.

*been fabricated or doctored; or the "one-shot stop" results have, or both have.*

This would be a valid point if the Strasbourg results really had a "near-perfect" correlation with the M&S OSS data. The actual correlation coefficient between the Strasbourg tests and the OSS data set depends on which of the data sets are used, the model, and the estimation technique. Steve Fuller reported R = 0.92 [MAS96 Ch28], and we found R = 0.87 [COC06b]. M&S report a Spearman Rank Correlation Coefficient between the Strasbourg data and their OSS rating as 0.89 [MAS96 p38]. These levels of correlation show substantial agreement without coming anywhere near the exaggerated claim that the level of agreement is "near perfect" or "too good to be true."[6]

*iv. Difficulty obtaining 600 goats at target weight*
Obtaining 600 animals within 2.5% of 160 lbs is a challenging task, but the practical difficulty in performing an experiment is not usually considered an indicator of fraud. In addition, without knowing the market conditions local to the experiment, one cannot ascertain the number of animals from which selection was possible.

Those familiar with animal husbandry know that it is possible to purchase goats in a much broader weight range (say 140-175 lbs) and bring most of these goats into the prescribed weight range within a reasonably short time (several weeks) with a carefully controlled diet and monitoring of their weight. With a carefully designed diet, adult livestock in this size range can easily gain 2-3 lbs per week or lose 1.5-2 lbs per week without a negative impact on overall health.

*E. Criticism of publication details*

*i. Anonymity of authors*
Anonymous authorship is the main reason given by the FBI committee for concluding that the Strasbourg report is a fraud [FAC94a]. In his summary of their opinion, Fackler asked, "Why else would experimental results be circulated anonymously?"

As researchers who ourselves have been involved with live animal research [COC06d, COC06e, COC07a], we can describe some reasons:
- *Avoiding harassment by animal rights fanatics*

We developed a method to study handgun bullets in deer [COC06d, COC06e]. In one year, our hunting efforts were disturbed by animal rights activists on a dozen occasions. We also operated a pasture-based livestock operation [CPD02, HOL05]. Livestock were let out of pastures on several occasions by animal rights activists, and one animal was "rescued" (stolen). One of the authors (MC) was threatened by an animal rights activist with a shotgun. Other deer management operations in the county were subjected to litigation by the animal rights activists.

Scientists connected to live animal research are frequently subjected to harassment and threats from animal rights activists and prefer to be discreet. This is widely recognized in the law enforcement, scientific and firearms communities (Editorial, American Rifleman, Dec 2005 p10):

*PETA[7] wants to stop medical advancements that use animal research in any way.*

and

*PETA's agenda is being forced upon society with acts of violence and terrorism . . . This eco-terrorism movement is so dangerous, the FBI has declared it America's No. 1 domestic terrorist threat on American soil . . . Scientists, doctors and their families are having property destroyed and are getting hate mail, letters loaded with razor blades and rat poison, and death threats and bomb threats.*

- *Nondisclosure agreement with funding source*

Even though the Strasbourg tests could have been legally conducted under the NAS guidelines, it is a borderline case that could result in litigation. An overzealous prosecutor, an activist judge, or an animal rights lawyer could create grief and expense for the scientists and funding source, even if they ultimately prevailed in court. The simplest and cheapest way to avoid potential legal battles is to maintain anonymity.

- *Avoiding difficulty at home institutions*

Most participating scientists in this kind of project maintain full time jobs in faculty or medical positions. Both the animal rights and the anti-gun interests at many educational and medical institutions can stir up trouble for participants in research projects that are not "politically correct." Most of the educational and medical community is unfriendly to firearms and live animal research that is perceived as "borderline."

- *Nondisclosure agreement with site and/or colleagues*

One site where our live animal research was conducted asked not to be identified, and several contributing/advising scientists asked to remain anonymous.

---

[6] Fackler also misquotes M&S as finding an "extremely high rank correlation" between Strasbourg and their own data. Quotation marks are usually reserved for exact quotes, but the actual words of M&S describing the correlation between Strasbourg and the OSS data set are "impressive" and "credible."

[7] PETA is the name of an activist animal rights group, People for the Ethical Treatment of Animals.

Does anonymous publication of scientific results imply fraud? Completely anonymous publication of scientific results is relatively rare, but there are notable historical exceptions. Copernicus delayed publication of his heliocentric model of the solar system until after his death to avoid persecution from the Inquisition. Under similar threat, Galileo published some of his results under a pseudonym. A precursor to Darwin's "Origin of the Species" entitled *Vestiges of the Natural History of Creation* was also published anonymously (eventually credited to Robert Chambers).[8]

Many pretend that they would boldly stand by their ideas in spite of fears of negative consequences, telling themselves they would not have feared the Inquisition and shrunk back in the manner of Galileo and Copernicus. But few men demonstrate the level of bravery we all hope to possess.

Who, when presented with a table of his "heretical" writings, would say to the Inquisition not only "yes these are mine" but also "and I have written others"? Who, facing fear of death and imprisonment would refuse to recant his writings and say bravely, "Here I stand, I cannot do otherwise"? While there are notably few men like the original Martin Luther [BAI50], there are many more men like Copernicus and Galileo who prefer to publish posthumously or anonymously, and even many more who simply remain silent.

Willingness to suffer negative legal, professional, and personal consequences for controversial work is admirable. However, it is not essential to validity. Maybe a modern day Martin Luther can criticize the Strasbourg authors for remaining anonymous, but we will not.

*ii.     Documentation not available*
It is frequently mentioned regarding the Strasbourg tests that experiments are not valid unless accompanied by video or photographic documentation. However, it is not true that this level of documentation always accompanies live animal research studies. Authors of live animal research studies are careful not to provide animal rights activists with photographic evidence that can be presented in court to obtain a restraining order or posted on a web site to enhance a funding campaign or ignite protests.

---

[8] United States history also has examples of important and influential works published anonymously; the Federalist Papers are the most notable example. In addition, Ben Franklin had a number of anonymous publications, including not only political but also scientific works. Should some of Ben Franklin's scientific contributions be considered fraudulent because they were published under a pseudonym?

The standard of documentation in any scientific study is a description of the methods and results sufficient for independent researchers to repeat the experiment. Photographs and diagrams need only accompany a report to the degree necessary to communicate the essential elements of the method and results. The existing Strasbourg report meets this standard without the inclusion of photographs. An independent research group could shoot similar goats with the same shot placement and observe whether the incapacitation times are consistent with the original report.

The burden of proof for proponents of a scientific theory is to provide evidence supporting that theory in the results of repeatable experiments. In the absence of governmental or institutional requirements mandating a specified level of documentation, the repeatability of an experiment (rather than the documentation) is generally considered to meet this burden of proof.

The burden of proof for those asserting a scientific fraud rests with those asserting fraud. It is fallacious for those asserting fraud to assume a posture of "guilty until proven innocent" in the absence of any eyewitness, documentary, or physical evidence of fraud. Shifting the burden to the original authors to prove themselves innocent of a fraud accusation is an example of the burden of proof fallacy.

*Conclusion on Strasbourg*
In the absence of support or direct contradiction from other experiments, the veracity of the Strasbourg tests should fairly be considered to be an open question. Neither the anonymity of the authors nor other criticisms offered are sufficient to consider the report fraudulent. Rather than lean too heavily on (possibly biased) expert opinions, the veracity of the report should be determined by the degree to which the reported results find support in other experimental findings.

The principle observations of the Strasbourg tests find support in the results of studies in pigs [SHS90a, SHS90b], dogs [WWZ04], rats [THG97, TLM05, and references therein], humans [MAS96, OBW94], whales [KNO03], and deer [COC06d]. The pressure wave hypothesis also receives direct experimental support in an experiment applying a ballistic pressure wave without a wound channel [COC07a]. In addition, traumatic brain injury has been linked to the ballistic pressure wave [COC07b].

**V.     Exaggerated criticisms of the Marshall and Sanow "one shot stop" data set**
The Marshall and Sanow (M&S) study [MAS92, MAS96] collected historical data for a large number of shootings with a wide variety of handgun loads in an attempt to quantify the relative effectiveness of different handgun loads. The selection criteria included events where a criminal attacker was hit in the thoracic cavity with a single bullet and where the bullet and load could be accurately identified.

Events where the attacker ceased the attack without striking another blow or firing another shot were classified as "one-shot stops." Events where the attacker delivered subsequent blows, fired subsequent shots, or ran more than 10 feet were classified as "failures." Events where the attacker retreated but covered less than 10 feet were classified as successful stops also. The "one-shot stop" percentage (OSS) was determined by dividing the total "stops" by the total number of events meeting the selection criteria (stops + failures).

Most epidemiological-type studies of this size and complexity represent compromises between the breadth of the selection criteria and the number of data points available. Other researchers might have made different choices about these trade-offs or differed in implementation details, but on the whole, the M&S OSS data set is a valuable contribution to understanding incapacitation via handgun bullets.

A number of criticisms have addressed the work of M&S [ROW92, FAC94a, FAC94b, FAC97a, FAC99a, FAC99b, MAC97, MAA99]. These vary in validity and implications and were considered carefully before we used the OSS data set to develop an empirical model of the ballistic pressure wave contributions to incapacitation [COC06b].

The published criticisms include unjustified *ad hominem* attacks and other rhetorical fallacies, gross exaggerations depending upon unjustified presuppositions, and valid concerns affecting the accuracy but not the validity of considering the OSS rating as a measure of relative handgun load effectiveness.

A.  Ad hominem attacks, appeal to ridicule fallacy, and bandwagon fallacy

The *ad hominem* attacks address both the authors (Marshall and Sanow) and those "lunatic fringe" who find value in the OSS data [FAC99a]. *Ad hominem* attacks, the bandwagon fallacy, and impugning the motives of authors without evidence have no place in the peer-reviewed scientific literature.

In their critical book review [ROW92] of the first stopping power book [MAS92], Roberts and Wolberg write:

*No clear thinking person should fall prey to this nonsense, but some individuals with no background in science or those too indolent or busy to do their own thinking could be misled. This book is a bad joke...*

and

*...this confusing text provides the reader with a schizophrenic mixture of material.*

In the article, "Undeniable Evidence," [FAC99b] Fackler combines the bandwagon fallacy with an appeal to ridicule:

*From the outset, those with training in statistics, those schooled in the scientific method, those with experience in scientific research, and even those laymen who do their own thinking, have believed that the "one-shot stop" statistics published by Marshall were not collected as claimed, but simply made up – fabricated.*

and

*...it is difficult to imagine many remaining "believers" in the Marshall-Sanow camp. Sadly, however, there will always be "believers"; primarily those with meager intellectual gifts and those so inherently indolent that they will accept the word of some "expert" for anything that would require the least amount of thought.*

In "Sanow Strikes (Out) Again" [MAC97], MacPherson writes about Steve Fuller's empirical curve fit and Ed Sanow's comments on it [SAN96].

*This is just like someone's claim to have seen a unicorn; they didn't, and it doesn't really matter whether they sincerely think they did or are simply lying about it.*

and

*The bottom line is that this is just another pathetic chapter in a very pathetic story.*

In a review of the second M&S book [MAS96], Fackler begins by impugning the motives of the authors with a circumstantial *ad hominem* attack [FAC97a]:[9]

*The authors of this book are gunwriters who have close ties to bullet companies that specialize in lightweight handgun bullets shot at higher than usual velocities. They have published numerous articles, and a previous book, Handgun Stopping Power—the Definitive Study, which are essentially unabashed advertisements for this type of bullet.*

---

[9] *A circumstantial ad hominem is a fallacy because a person's interests and circumstances have no bearing on the truth or falsity of the claim being made.*
(*www.nizkor.org/features/fallacies/circumstantial-ad-hominem.htm*, accessed 11/4/2006).

One willing to commit the *to quoque* (you too!) fallacy could reply about Fackler:

*The author of this critique is closely tied to an FBI committee who selected a Winchester 147 grain JHP load for 9mm use and frequently advocates other heavy and slow bullets. He has published numerous articles, which are essentially unabashed advertisements for this kind of bullet.*

This reply is also fallacious, because it mistakenly frames the debate in terms of "light and fast" vs. "heavy and slow." Several "heavy and slow" bullets make a good showing in the M&S OSS figures, and several "light and fast" bullets perform poorly. For many cartridges, the top ranked loads are from Remington and Federal, two of the largest ammunition suppliers who do not "specialize in lightweight handgun bullets shot at higher than usual velocities" but rather offer an array of ammunition choices covering both the "light and fast" as well as the "heavy and slow" ends of the spectrum.

In addition, the M&S text goes into detail on the importance of bullet design beyond the oversimplification of considering only weight and velocity [MAS96 Ch 17-23].

There is a lot of information regarding bullet designs in the Stopping Power books, but no one has shown that Marshall and Sanow received compensation from ammunition companies in a time frame consistent with it swaying their OSS results.[10] Some of the leading bullet designers were willing to share details of the bullet design and development process, and it appears the authors included this information because of significant reader interest.

In this book review Fackler also writes:

*Fortunately, the great majority of law enforcement groups have ignored the Marshall and Sanow "Definitive Study" and opted for the heavier, slower bullets, which have proved far more reliable than the faster, lighter bullets they replaced.*

In addition to being the bandwagon fallacy, this argument is somewhat circular. Many law enforcement agencies simply follow the lead and standards of the FBI that Fackler had significant input in developing. Even if it were true, the fact that many law enforcement agencies are following the lead of the side of a debate that is considerably influential in law enforcement (because it is the FBI) is not compelling that experiments supporting the other side of a debate are invalid.

Ammunition selection by law enforcement agencies swings back and forth between "slow and heavy" and "light and fast," and these changes are often due to the vivid appeal fallacy rather than sound scientific reasoning based on statistical outcomes. Since handguns are inherently unreliable at creating rapid incapacitation [PAT89], there will be some vivid demonstrations of failures with any load that is selected. A "light and fast" 115 grain 9mm bullet failed to adequately penetrate in a famous 1986 gunfight in which two FBI agents lost their lives. This inspired wide adoption of a "slow and heavy" 147 grain JHP at 990 FPS.

Failures of the 147 grain JHP at 990 FPS in 9mm has lead to development of the .357 Sig cartridge and a number of agencies adopting a 125 grain bullet at 1350 FPS (light and fast). In the absence of sound statistical reasoning based on evaluation of a large number of shooting events and other sound scientific results, these decisions represent either the vivid appeal or expert opinion fallacies.

Fackler's book review [FAC97a] also includes the attack:

*... the authors of "Street Stoppers" fail to understand the most basic of scientific principles and discourse...*

Marshall and Sanow have more law enforcement than science backgrounds, and their studies and writing style reflect this. However, their general approach of designing an epidemiological type of study with selection criteria for inclusion and criteria for determining success or failure is scientifically sound. They formulated a useful methodology for addressing an important law enforcement issue and set about carefully collecting and analyzing data to address the issue.

Fackler's conclusion continues with the ridiculing:

*No intelligent reader will tolerate a nonfiction book devoid of references.*

apparently implying that interspersing references in the text (rather than collecting them in a bibliography) is unacceptable. Other authors in the field [MAC94] use the same approach yet garnered Fackler's praise and recommendation.

In the final paragraph of the book review, Fackler writes:

*"Street Stoppers" is a compilation of fantasy: written in the arrogant, dead certain tone of the con man. Everything echoes "trust me." The reader is constantly preached to, with evangelistic fervor: and without equivocation implored to believe in nonsense with no basis in established fact. This*

---

[10] Evan Marshall opened a gun shop in 2005 and can reasonably be presumed to have a current interest in selling ammunition.

*book is the antithesis of honest, intelligent, scientific discourse in which the evidence is laid out, dispassionately...*

The reasoning of M&S might not always be perfectly clear, and the content is not always perfectly organized, but their books are more dispassionately written and contain fewer *ad hominem* attacks (none) than the criticism offered in response.[11]

*B. Criticisms of experimental design*
There are several criticisms of the experimental design used by M&S to generate their OSS index.

*i. Excludes multiple hits (overly narrow selection criteria)*
As described above, the M&S OSS data set only includes events where the person receives a single bullet to the thoracic cavity. Those criticizing this criterion [FAC92a, MAA99] assert that cases where an attacker is shot multiple times should be included as failures. This position depends on the unsupported premise that attackers only sustain multiple gunshot injuries in cases where the first hit is ineffective.

Most law enforcement officers and many armed citizens are trained to fire multiple shots without stopping to assess whether the first shot has created the desired incapacitating effect.[12] Most law enforcement officers and many armed citizens train to a level of proficiency where firing multiple shots is possible in the time required for an attacker to fall down even if the first shot severs the spine (about 0.4 seconds). Since shooters are usually not re-evaluating to determine whether a second shot is necessary, the presence of a second hit cannot be taken as evidence of a first hit failure.

In the same way that a study of the effectiveness of a certain medicine needs to specify the dose of the drug, a study of bullet effectiveness needs to specify the "dosage" of a given load. Excluding multiple hit events from the data set is analogous to excluding walks from a batting average.

*ii. Thoracic cavity is too large (overly broad selection criterion)*
It has been suggested [ROW92, FAC97a] that any valid study of bullet effectiveness would have to precisely specify the anatomic structures impacted by the bullet. This assertion ignores the power of statistical approaches to quantify probabilistic outcomes given certain (even if broad) input information. For example, the idea behind actuarial tables is that a person's life expectancy can be predicted from relatively broad selection criteria such as a person's weight, age, income, marital status, drinking and smoking habits.

Narrowing the selection criteria allows for more accurate statistical descriptions given the same number of available data points, but the tradeoff is that narrowing the selection criteria reduces the included data points in an available data set. The designers of all epidemiological-type studies face this tradeoff, and setting the criteria at a given level does not impact the validity of a study. (It affects the accuracy of the conclusions.)

In fact, epidemiological data with broad selection criteria have been used to draw well-founded conclusions in wound ballistics. For example, Fackler correlates the declining incidence of specific infections with the increasing use of antibiotics on the battlefield to infer that "the benefits of systemic antibiotic usage have been incorrectly attributed to wound debridement" [FAC87a].

*iii. Too few data points*
Some loads included in the study have as few as 10 data points. Other loads have thousands. As long as the basic relationship between number of data points and accuracy of the OSS rating is considered in interpreting the results, this is not a problem. The M&S OSS data set grew to eventually include over 20,000 shooting events [MAS01].

*iv. Not a random sample*
The OSS data set does not represent a truly random sample [MAA99, ROW92], but depends on voluntary submission of shooting event reports to the authors of the study by law enforcement agencies. Thus the possibility exists for bias and falsification if the sources submitting the data exclude or falsify information because it does not fit their pre-conceived notions. However, crime statistics are often the result of voluntary reporting by law enforcement agencies. This might affect the accuracy of such statistics but does not make them completely useless.

Since the data set was compiled over more than a decade, it is also likely to include other non-random effects such as changes in officer training, improvements in product quality, and changing geographical distribution of the data set. For example, data from the warmer climates are less likely to represent the effects of heavy clothing than data from colder climates.

*v. Conclusion on M&S design*
In conclusion, the experimental design leading the M&S OSS rating is valid for considering relative effectiveness of different handgun loads. It is limited in accuracy primarily by different loads having different shot placement distributions within the thoracic cavity, by the number of data points available for each load, by voluntary submission of shooting reports, and by difficulties with completely objective determination of whether or not a shooting event resulted in a stop. Other factors impacting the accuracy of the one shot stop index are different loads having different distributions of impact conditions such as range and intermediate barriers and the potential for different loads to have different distributions of why subsequent hits were achieved.

---

[11] Evan Marshall and Ed Sanow have shown significant character and remarkable restraint in not answering their critics in kind.
[12] Exhortations to use the "double-tap" (two shots) or "shoot to lockback" are examples of this kind of training.

*C. Criticisms with analysis*

*i. Neurogenic shock*

Thinking a shock effect might be demonstrated in their data set, M&S invited Dennis Tobin, MD to write a section entitled "A Neurologist's View of Stopping Power" [MAS92 Ch2].

In their critique, Roberts and Wolberg [ROW92] write:

*The absurd claims made in this chapter…regarding the ability of a bullet to remotely stress and shock the central nervous system (CNS) are grossly inaccurate and completely unsupported by any scientific evidence… This pseudoscientific speculation is presented as verified scientific fact, yet allusion to Goransson's discredited data is the only scientific reference the authors cite to support their unjustified claims...*

*A thorough review of the scientific literature relating to wound ballistics has failed to identify any valid research papers which demonstrate that projectiles can exert a remote effect on the CNS.*

The remote CNS effect that Dr. Tobin suggests might fairly be classified as a hypothesis at the time of writing. However, referring to it as "pseudoscientific speculation" is exaggerated given his reference [GIK88]. The Goransson paper [GIK88] acknowledges that remote effects on the CNS were previously unexplored, but it gives a clear chain of references documenting peripheral nerve damage without a direct hit [LDL45, PGM46] and a paper describing the effect of pressure waves on compound action potentials in nerves [WES82]. It had also been previously documented [TCR82] that projectile wounds to the thigh produced pressure waves in the abdomen.

The Suneson *et al.* studies of remote pressure wave effects on the CNS may not have been available to Tobin when he wrote the original article. These studies [SHS90a, SHS90b] appear to be the first studies to fully document remote pressure wave effects on the CNS. These papers were published more than two years before the critique of Roberts and Wolberg [ROW92]. Therefore, the assertion that a thorough review of the literature missed these papers is surprising, especially since a comment on these papers was published in the widely read **Wound Ballistics Review** [FAC91a] to which both of the critical authors had been contributors.

The allusion to Goransson's data [GIK88] as "discredited" is also an unsupported exaggeration. The data of Goransson *et al.* find substantial support in the work of Suneson *et al.* [SHS90a, SHS90b] and Wang *et al.* [WWZ04]. Roberts and Wolberg fail to cite any articles critical of this work. Note that in the same article criticizing M&S for not adequately documenting their work, Roberts and Wolberg failed to document their claim that Goransson's work had been discredited.[13]

*ii. Pseudo-scientific formulas*

In Chapter 17 of **Handgun Stopping Power** [MAS92], M&S present the results of linear least squares fits between their OSS rankings and parameters commonly measured in ballistic gelatin. They find that the OSS rating correlates better with temporary stretch cavity (R = 0.8) than permanent crush cavity (R = 0.6). This contradicts the view that only permanent cavitation contributes significantly to wound trauma incapacitation [MAC94, PAT89]. Roberts and Wolberg write in criticism [ROW92]:

*The pseudoscientific formulas purported to predict "stopping power" which are presented in Chapter 17 are unsupported by any scientific evidence. These formulas appear to be completely meaningless since they are based on the irrelevant and misleading "street results" and since the gelatin tests used by the authors appear to be flawed and inaccurate.*

The data set is based on a sound design. The criticism of the M&S gelatin testing will be considered below. Using regression analysis and correlation coefficients to determine whether two data sets are related is a time-proven scientific approach. Therefore, there is little room for valid criticism here[14] beyond the validity of the OSS data set itself and the validity of the criticism of the gelatin measurements.

*iii. Gelatin results inaccurate*

Roberts and Wolberg write [ROW92]:

---

[13] Usually a negative comment "discrediting" an article would appear in the same journal as the original paper. No such criticism of Goransson *et al.* appears in *J Trauma*. Using commonly available scholarly search engines, one finds several peer-reviewed articles that cite Goransson *et al.* favorably, but none that "discredit" the data.

[14] A model with proper limiting behavior might be preferred [COC06b].

*The penetration depths of test shots into ordnance gelatin listed by Marshall and Sanow in Table 17-1 [MAS92] are consistently deeper than those reported by other wound ballistic research facilities throughout the United States, as illustrated by the following [three] examples...*

Comparing all 24 loads in the table with independently produced results would be a more sound method of demonstrating systematic errors in the gelatin data. The reader is left uncertain whether the three referenced examples are typical or worst cases.

Even if the M&S gelatin results systematically overestimate penetration, this does not nullify their correlations. Accuracy in many scientific fields gradually increases over time due to adoption of better testing standards. The gradual adoption of carefully calibrated 10% gelatin at a controlled temperature in the 1980's and 1990's improved evaluation of bullet penetration and expansion. However, the suggestion that M&S used uncalibrated gelatin does not nullify their results. It suggests only that their gelatin measurements might not be optimally accurate.

### D. Criticisms of experimental implementation

#### i. Better with time

Many of the reported OSS ratings increase slightly over time. This has been criticized [MAA99, FAC97a] using the logic that if new shootings were simply added to the data set, "about as many "one-shot stop" percentages would be expected to decrease as would be expected to increase" [FAC97a]. Rather than being an indication of "too good to be true," which Fackler asserts implies fraud, these slight increases in 48 of 60 loads can be understood to result from a variety of non-random factors, including improvements in loads meeting velocity specifications, improved training, more data from warmer climates, etc. The premise of randomness that creates the expectation of sameness over time is flawed.

#### ii. Data added too fast

van Maanen [MAA99] points out that the published data set increases from 6136 events in 1992 [MAS92] to 20742 events in 1996 [MAS96]. He incorrectly infers

*…the number of cases added to the Marshall and Sanow "data base" was 14606 between 1992 and 1996; an average of 10 cases per day 365 days a year.*

van Maanen goes on to comment that because many cases do not meet the selection criteria, it would have been prohibitively time consuming for the authors to carefully evaluate the large number of cases to determine whether and how they should be included in the data set, writing:

*One must conclude that the care taken in evaluating the shootings that make up the Marshall and Sanow "data" is at best far less than the authors' claims in the description of the process. One could easily conclude that the time available for evaluating each case inherently makes the integrity of any evaluation process indistinguishable from simply making up the cases outright.*

This analysis on the rate of new shootings added to the data set is in error because the book published in 1992 did not contain any data for .22LR and .25 ACP loadings, which account for 10288 events in the data published in 1996. Most of these events had already been collected and analyzed in 1992, but M&S chose not to publish them at that time.

Thus, the number of events added to the data set was much smaller than the number of additional events published in 1996. The average number of cases added to the data set per day was probably less than five.

van Maanen's analysis also employs the unlikely premise that the entire labor burden of shooting analysis was personally on the authors. Using trained assistants to perform "triage" on shooting reports would have greatly streamlined the analysis process for the principal investigators. Shootings clearly not meeting inclusion criteria could have been sorted by trained assistants to save time for the principal investigators.

The claim that the principal investigators did not have the time to personally perform all of the described procedures is not usually considered a valid reason to conclude that results are indistinguishable from outright falsified data.

#### iii. Events removed from data set

Analysis [MAA99] of the data set over time shows that shooting events for 16 loads must have been removed from the data set between 1992 [MAS92] and 1996 [MAS96]. Rather than considering possible valid reasons for removing events, van Maanen takes this to support his prior conclusion of fraud:

*Any kind of secret reevaluation of the results is a clear-cut violation of doing research because it is one form of "fudging" the data.*

This ignores other reasonable possibilities:
- Discovery and correction of clerical errors resulting in double counting.
- Discovery that some data sources proved unreliable and removal of the affected subset of data.

Optimally, M&S would have provided explanation and justification for these adjustments. However, given attacks and fallacious rhetoric that had already appeared, they might have been reasonably reluctant to provide critics with ammunition regarding the validity or accuracy of their first book [MAS92]. Finally, it should be noted that less than 20% of the loads studied indicate removal of events.

*iv. A source says reports fail to meet selection criteria*
In addition to presenting the OSS data set and method used to produce it, the M&S books [MAS92, MAS96] present other material of reader interest including background on new bullet designs, results of shooting ballistic gelatin, tactical advice, physiological information, and a number of shooting event anecdotes written in a journalistic style intended to capture reader interest rather than document the OSS data.

Roberts and Wolberg suggest fraud [ROW92a] because some of the anecdotal shooting reports M&S ascribe to Wolberg do not meet criteria for inclusion in the OSS data set (insufficient documentation). However, there is no indication that these events are included in the data set [MAS92, p121]. A significant fraction of the journalistic-type shooting reports clearly do not meet the criteria for inclusion. Thus the conclusion of Roberts and Wolberg that the data set is unreliable is fallacious. It rests on the false premise that all the anecdotal shooting reports were included in the data set. In the second book, M&S clearly state that many of the anecdotal accounts are not included in the data set [MAS96, p19].

*E. Criticisms of publication details: source data is not shared*
M&S have been criticized for not publishing or sharing their raw source data with other investigators [FAC97a], even though they provide voluminous data in aggregate form [MAS92, MAS96, MAS01]. The criticism implies that sharing raw data is a universally accepted scientific standard necessary to establish the validity of a work.

This is not true. In the absence of legal or institutional requirements governing the availability of raw data, investigators have the sole right to make decisions regarding the sharing of raw data and other proprietary research products (computer programs, designs, photographs, etc.).

It is an exaggeration to conclude that researchers who limit distribution of raw data are hiding something. Valid reasons to restrict access to raw data and documentation include:

1. Protecting the privacy of sources.
2. Maintaining non-disclosure agreements.
3. Maintaining individual patient/victim confidentiality.
4. Maintaining competitive advantage on grant proposals, future discoveries, and new product development.
5. Protecting investment acquiring the raw data.
6. Avoiding costs incurred in sharing the data.
7. Avoiding misuse/misinterpretation leading to legal liabilities.

In the absence of legal or institutional requirements, access to the raw data, computer programs, documentation, or other proprietary products of scientific research is at the discretion of the investigators. Scientists are often willing to share with other scientists who have developed a trust relationship and offer possibility for mutual benefit from the sharing relationship. Benefits that commonly induce researchers to share proprietary data or other research products include:

1. Hope of discoveries leading to jointly authored publications unlikely to occur otherwise.
2. Possibility of securing additional funding.
3. Acknowledgements in peer-reviewed publications.
4. Greater awareness of the scientist's contributions.
5. Maximal use made of live animal studies.
6. Overt *quid pro quo* exchanges of one research product for another (or for money).

Trust, goodwill, and mutual admiration are the bedrock of this kind of scientific collaboration, so it is not surprising that Marshall and Sanow declined to share their raw data and documentation with those who had been antagonistic. In contrast, Marshall and Sanow have offered to share their data with some authors [AYO07].

It is also noteworthy that referencing unpublished notes and data is common in the field of wound ballistics. The highly esteemed Martin L. Fackler references unpublished data in several papers [FAC96a, FAC87a, FSM84a]. Dr. Carroll Peters, to whom Fackler referred to as a "renown academic" refers to unpublished data, personal communications, and notes [PET90a].

F. *Conclusion on Marshall and Sanow*
Answering every fallacious and exaggerated criticism would take a lot more space and involve an unreasonable level of quibbling over details that are not important to the validity of the OSS rating as an indicator of relative bullet effectiveness.

The weaknesses in the M&S OSS ratings affect the accuracy as a relative measure of stopping power, but not the overall validity. The accuracy of the OSS assignment for different loads depends on the details of the given load but is most strongly dependent on the number of data points included for that load. For most loads containing N events, the accuracy of the OSS rating is probably within a factor of two of the statistical uncertainty, $0.5/N^{1/2}$. This level of accuracy is sufficient to provide confirmation of the pressure wave contribution to rapid incapacitation.

The correlation between the M&S OSS ratings and the Strasbourg average incapacitation times [COC06b, MAS96] is also compelling support. Add this to the results that the M&S OSS ratings can be accurately modeled as a function of pressure wave magnitude [COC06b] and that a growing body of literature [CHA66, TCR82, SHS90a, SHS90b, OBW94, WWZ04, THG97, TLM05, COC07a, COC07b] supports connections between pressure wave, injury, and incapacitation. The criticisms of the M&S OSS data set are exaggerated, and the M&S data set has considerable consistency with results of independent work.

## VI.     Overall Conclusion

In light of the demonstrated *ad hominem* attacks, exaggerations and fallacies in the criticisms of these experimental findings, one wonders whether the critical authors were depending more on their reputation as experts and the quantity of their fallacies (*ad nauseum* fallacy) rather than quality arguments, sound reasoning, and repeatable experiments. The critical authors left quite a paper trail in the literature, but reason, the scientific method, and repeatable experiments and analysis have shown the original works to be more sound than the published criticisms.

This is not to say that the criticized works are perfect or beyond reasonable criticism or scrutiny, only that the ultimate validity and value of these works should be determined by the whole body of relevant literature, as well as future work.

The level of exaggeration, *ad hominem* attack, and other rhetorical fallacies in these critical reviews far exceeds the boundaries of sober and rational discourse that characterize modern peer-reviewed scientific literature. Even in areas rife with pseudo-science (such as nutritional supplements, psychic phenomena, global warming, the creation/evolution debate), the tenor of the peer-reviewed scientific literature is remarkably more restrained and sound in its reasoning.

Since we (the authors, Michael Courtney and Amy Courtney) are now contributors in a field with unusually high levels of "ammonia and acetic acid," we would like to express our sincere hope that future debate will be characterized by more civilized discussion without resorting to personal attacks, insults, and unrestrained rhetorical fallacies. Going beyond the accepted boundaries of scientific discourse reflects poorly on the field, on the law enforcement interests in the discussion, and on firearms-related issues in general.

[SBC01] Sokolov DL, Bailey MR, Crum LA: Use of a Dual-Pulse Lithotriptor to Generate a Localized and Intensified Cavitation Field. Journal of the Acoustical Society of America 110(3):1685-1695, 2001.

[SHA02] Shaw NA: The Neurophysiology of Concussion. Progress in Neurobiology 67:281-344; 2002.

[SHK90] Suneson A, Hansson HA, Kjellström BT, Lycke E, and Seeman T: Pressure Waves by High Energy Missile Impair Respiration of Cultured Dorsal Root Ganglion Cells. The Journal of Trauma 30(4):484-488; 1990.

[SHL89] Suneson A, Hansson HA, Lycke E: Pressure Wave Injuries to Rat Dorsal Cell Ganglion Root Cells in Culture Caused by High Energy Missiles, The Journal of Trauma. 29(1):10-18; 1989.

[SHS87] Suneson A, Hansson HA, Seeman T: Peripheral High-Energy Missile Hits Cause Pressure Changes and Damage to the Nervous System: Experimental Studies on Pigs. The Journal of Trauma. 27(7):782-789; 1987.

[SHS88] Suneson A, Hansson HA, Seeman T: Central and Peripheral Nervous Damage Following High-Energy Missile Wounds in the Thigh. The Journal of Trauma. 28(1 Supplement):S197-S203; January 1988.

[SHS90a] Suneson A, Hansson HA, Seeman T: Pressure Wave Injuries to the Nervous System Caused by High Energy Missile Extremity Impact: Part I. Local and Distant Effects on the Peripheral Nervous System. A Light and Electron Microscopic Study on Pigs. The Journal of Trauma. 30(3):281-294; 1990.

[SHS90b] Suneson A, Hansson HA, Seeman T: Pressure Wave Injuries to the Nervous System Caused by High Energy Missile extremity Impact: Part II. Distant Effects on the Central Nervous System. A Light and Electron Microscopic Study on Pigs. The Journal of Trauma. 30(3):295-306; 1990.

[STR93] The Strasbourg Tests, presented at the 1993 ASLET International Training Conference, Reno, Nevada.

[THG97] Toth Z, Hollrigel G, Gorcs T, and Soltesz I: Instantaneous Perturbation of Dentate Interneuronal Networks by a Pressure Wave Transient Delivered to the Neocortex. The Journal of Neuroscience 17(7);8106-8117; 1997.

[TCR82] Tikka S, Cederberg A, Rokkanen P: Remote effects of pressure waves in missile trauma: the intra-abdominal pressure changes in anaesthetized pigs wounded in one thigh. Acta Chir. Scand. Suppl. 508: 167-173, 1982.

[TLM05] Thompson HJ, Lifshitz J, Marklund N, Grady MS, Graham DI, Hovda DA, McIntosh TK: Lateral Fluid Percussion Brain Injury: A 15-Year Review and Evaluation. Journal of Neurotrauma 22(1):42-75; 2005.

[WES82] Wehner HD, Sellier K: Compound action potentials in the peripheral nerve induced by shockwaves. Acta Chir. Scand. Suppl. 508: 179, 1982.

[WWZ04] Wang Q, Wang Z, Zhu P, Jiang J: Alterations of the Myelin Basic Protein and Ultrastructure in the Limbic System and the Early Stage of Trauma-Related Stress Disorder in Dogs. The Journal of Trauma. 56(3):604-610; 2004.

[WOL91] Wolberg EJ: Performance of the Winchester 9mm 147 Grain Subsonic Jacketed Hollow Point Bullet in Human Tissue and Tissue Simulant. Wound Ballistics Review. Winter 91:10-13; 1991.
**About the Authors**

*Amy Courtney* currently serves on the faculty of the United States Military Academy at West Point. She earned a MS in Biomedical Engineering from Harvard University and a PhD in Medical Engineering and Medical Physics from a joint Harvard/MIT program. She has taught Anatomy and Physiology as well as Physics. She has served as a research scientist at the Cleveland Clinic and Western Carolina University, as well as on the Biomedical Engineering faculty of The Ohio State University.

*Michael Courtney* earned a PhD in experimental Physics from the Massachusetts Institute of Technology. He has served as the Director of the Forensic Science Program at Western Carolina University and also been a Physics Professor, teaching Physics, Statistics, and Forensic Science. Michael and his wife, Amy, founded the Ballistics Testing Group in 2001 to study incapacitation ballistics and the reconstruction of shooting events. www.ballisticstestinggroup.orgRevision information:
*13 December 2006 to 31 July 2007:* Fixed typographical errors. Updated references, contact information, and biographies. Added references [CHA66], [AYO07], [FSM84a], [PET90a], [LDL45], [PGM46], [MYR88], [WES82], [TCR82], [COC07a] and [COC07b].

18